\documentclass[]{article}%

\usepackage{graphicx}%
\usepackage{multirow}%
\usepackage{amsmath,amssymb,amsfonts}%
\usepackage{amsthm}%
\usepackage{mathrsfs}%
\usepackage[title]{appendix}%
\usepackage{xcolor}%
\usepackage{textcomp}%
\usepackage{manyfoot}%
\usepackage{booktabs}%
\usepackage{algorithm}%
\usepackage{algorithmicx}%
\usepackage{algpseudocode}%
\usepackage{listings}%
\usepackage{setspace}
\usepackage{lineno}
\usepackage{placeins}
\usepackage{pdfpages}
\usepackage{url}
\usepackage{authblk}

\begin{document}

\title{Urban Sensing Using Existing Fiber-Optic Networks}

\author[1]{Jingxiao Liu\thanks{jingxiao@mit.edu}}
\author[2]{Haipeng Li}
\author[3,a]{Hae Young Noh}
\author[1,4,a]{Paolo Santi}
\author[2,a]{Biondo Biondi}
\author[1,5,a]{Carlo Ratti}

\affil[1]{Senseable City Laboratory, Massachusetts Institute of Technology, Cambridge, MA 02139, USA}
\affil[2]{Department of Geophysics, Stanford University, Stanford, CA 94305, USA}
\affil[3]{Department of Civil and Environmental Engineering, Stanford University, Stanford, CA 94305, USA}
\affil[4]{Istituto di Informatica e Telematica del CNR, Pisa, Italy}
\affil[5]{Dipartimento ABC, Politecnico di Milano, Italy}
\affil[a]{These authors jointly supervised this work}

\maketitle

\abstract{
\doublespacing
The analysis of urban seismic signals offers valuable insights into urban environments and society.
Yet, accurate detection and localization of seismic sources on a city-wide scale with conventional seismographic network is unavailable due to the prohibitive costs of ultra-dense seismic arrays required for imaging high-frequency anthropogenic sources.
Here, we leverage existing fiber-optic networks as a distributed acoustic sensing system to accurately locate urban seismic sources and estimate how their intensity varies over time.
By repurposing a 50-kilometer telecommunication fiber into an ultra-dense seismic array, we generate spatiotemporal maps of seismic source power (SSP) across San Jose, California.
Our approach overcomes the proximity limitations of urban seismic sensing, enabling accurate localization of remote seismic sources generated by urban activities, such as traffic, construction, and school operations. 
We also show strong correlations between SSP values and environmental noise levels, as well as various persistent urban features, including land use patterns and demographics. 
}

\section*{Introduction}
\par
Cities, as epicenters of human activity and major contributors to global emissions, have drawn increased research interests aimed at improving urban policies, enhancing the quality of life, and promoting sustainability~\cite{TRINDADE20171}. 
Urban science employs quantitative and modeling approaches from various disciplines to achieve these goals~\cite{acuto2018building, bettencourt2021introduction}. 
Central to this field is the availability of large-scale datasets capturing various urban ``signals'' with precise spatial and temporal resolution~\cite{doi:10.1098/rsta.2016.0115, Ilieva2018}. 
Despite advances in urban sensing and big data approaches~\cite{dong2024defining, KANDT2021102992, rs15051307}, comprehensive urban data access remains limited due to issues, including data ownership, privacy concerns, and high costs~\cite{9080609, 10.1145/1411759.1411763}.
Recently, urban seismic signals have been used to monitor urban environments cost-effectively and pervasively. 
These signals carry valuable information for characterizing urban dynamic features such as environmental conditions, traffic patterns, and cultural and social activities~\cite{https://doi.org/10.1029/2020GL089931,urban1, groos, https://doi.org/10.1029/2023JB028033, doi:10.1126/science.abd2438}. 
However, seismic signals related to urban activities, particularly those at frequencies above 1 Hz, are subject to scattering and attenuation over short distances~\cite{https://doi.org/10.1002/2015GL063558}. 
Consequently, accurately estimating the spatiotemporal distribution and intensity of urban seismic sources, a process known as seismic source mapping, requires ultra-dense seismic arrays, which are not feasible to install in urban areas due to high costs and disruptions. 
Hence, it is unclear how these urban seismic sources are distributed within the city and how they are related to dynamic and persistent urban features.
\par
Here, we explore the use of existing and ubiquitous urban infrastructure -- telecommunication optical fibers -- to map urban seismic sources. 
By repurposing existing fiber-optic cables into a sensing network using distributed acoustic sensing (DAS)~\cite{2021AREPS..49..309L, 10.1785/0220190112}, we create an ultra-dense and cost-effective seismic array capable of covering extensive urban areas.
Integrating seamlessly with the existing telecommunication infrastructure, our approach provides a scalable solution for continuously recording urban seismic signals.
While DAS has been used to detect and locate distant natural seismic sources like earthquakes~\cite{li2023break, se-12-915-2021} and ocean waves~\cite{doi:10.1126/science.aay5881, Fernandez-Ruiz:22}, its applications in urban science~\cite{10.1145/3596262, se-12-219-2021,Jiajing:19,wang2021ground}, such as traffic and footstep detection, have been limited by proximity sensing — they have been used to monitor ground motions and quasi-static deformations due to activities that occur close to the sensors, such as traffic along the road on which the fiber is laid.  
Moreover, existing DAS applications for urban sensing have been developed for specific purposes, like traffic estimation or bridge health monitoring~\cite{liu2023turning,doi:10.1177/14759217241231995}, and are not adaptable for general-purpose urban sensing.
\par
In this study, we integrate seismic interferometry and beamforming algorithms to overcome the proximity limitations of previous urban seismic sensing approaches. 
Our method models the propagation of seismic surface waves to estimate the spatiotemporal distribution of seismic source power (SSP), defined as the average seismic energy per unit of time within an area. 
Our approach achieves precise localization and estimation of various seismic sources occurring remotely from optical fibers on a two-dimensional spatial urban scale.
We validate our seismic source mapping results against various urban seismic activities with known coordinates and timings, including city-wide traffic movements and human activities at construction sites and schools. 
Additionally, we find that SSP intensity is influenced by and significantly correlated with several persistent urban features, such as land use patterns, average daily traffic, points of interest density, and demographics.

\section*{Results}
\subsection*{Scalable seismic source mapping}
\par
The continuous seismic monitoring capabilities of DAS using existing telecommunication fiber networks present a unique Large-N array~\cite{clayton2011community} to capture seismic signals from various urban seismic sources. 
The recorded seismic signals are analyzed to estimate spatiotemporal maps of SSP (Fig. 1). 
This analysis utilizes seismic data generated from passive sources and collected through existing, unused fiber-optic cables, known as dark fibers. 
Therefore, it is highly scalable and integrates seamlessly with the existing telecommunication infrastructure.
By connecting an interrogator unit to one end of a 50-km long dark fiber in San Jose, California, we created 50,000 virtual dynamic-strain sensors (DAS channels) continuously sampled at 200 Hz and distributed in one-meter spacing along the fiber, mainly along the roadways.
These virtual sensors, each costing the equivalent of a few dollars when splitting the cost of the interrogator unit, operate without interrupting telecommunication services or necessitating on-site sensor installation and maintenance~\cite{10.1785/0220190112, 2021AREPS..49..309L}.
We continuously recorded seismic signals from the fiber-optic cables encircling three key regions over six days from September 20th to 25th, 2023. The dataset is detailed in Methods section.
\par
Our method begins by using vehicle-induced seismic signals to estimate surface wave velocities.
This estimation, combined with the beamforming technique~\cite{pillai2012array,10.1155/2012/292695,7342886,van2021evaluating}, enables the mapping of other urban seismic sources (Methods and Fig. 1).
In particular, to estimate surface wave velocities, we utilize a specialized Kalman filter algorithm to select vehicle-induced surface wave windows~\cite{10.1145/3596262} (Methods and Supplementary Fig. 1), followed by seismic interferometry on these windows. 
Leveraging the estimated locations of vehicles, we construct the virtual shot gathers (Methods and Supplementary Fig. 2) with enhanced signal-to-noise ratios and reduced computational costs~\cite{https://doi.org/10.1029/2023JB028033} when compared to standard ambient noise interferometry.
Moreover, to accurately estimate the seismic source power, we evaluate the attenuation in near-surface structures of each region by calculating a frequency-independent quality factor (Methods and Supplementary Fig. 6). 
Then, we implement frequency domain delay-and-sum beamforming to create spatiotemporal maps of SSP.
The beamforming method aligns and sums DAS signals in the frequency domain to determine source location and power using the seismic waves’ travel time delays, which are calculated using the distances from the source to the DAS channels and the estimated surface wave velocities.
Detailed descriptions of our methodology are outlined in Methods section.

\subsection*{Validations of seismic source mapping}
\par
The results of seismic source mapping have been validated against a diverse set of urban seismic events with precisely known coordinates and timings.
To establish ground truth, we use the fiber co-linear with roadways and railways to accurately estimate the positions and timings of seismic sources, such as trains and trucks moving on these roads and rails.
The self-weight of moving vehicles causes the ground to deform elastically, producing quasi-static DAS signals~\cite{10.1145/3596262}. 
Locations of the trucks and trains are estimated by detecting peaks in these quasi-static signals of the co-linear fiber responses.
\par
We then use DAS data only from the remote fibers to map these urban seismic sources.
In particular, the first segment of the telecommunications cable (DAS channels 0 to 3,500) in Region 3, as shown in Fig. 2a, runs adjacent to a railway track. 
The passing of the freight train along this track generates strong seismic waves, detectable within an aperture of 2 kilometers centered on the train position (Supplementary Fig. 5a and d). 
Similarly, seismic waves induced by truck movements can be identified from as far as 1.5 kilometers away (Supplementary Fig. 5b and d). 
Another significant example is the seismic signals generated by construction activities. Public records of development projects in San Jose, California, indicate a construction site near DAS channel 7300. 
Seismic waves generated by construction activities are detectable by DAS channels from up to 500 meters (Supplementary Fig. 5c and d).
\par
The agreement between the predicted and actual arrival times of surface waves generated by the aforementioned events assesses the accuracy of our surface wave velocity estimations (Fig. 2a-c).
Following the frequency-domain beamforming, spatial maps of SSP for these events were constructed (Fig. 2d-f and Supplementary Fig. 8). 
Importantly, to validate seismic source mapping using remote optical fibers, we utilize only DAS channels located far from the train, truck, and construction sources for beamforming. 
The alignment between the mapped seismic source hotspots and the actual event locations demonstrates the effectiveness of our approach in both detecting events and accurately estimating their locations.
The mapping of SSP at the construction site further verified the capability of our approach in estimating stationary seismic sources.
\par
We extend the validation of our seismic source mapping over the six-day period. 
The six-day dataset is divided into 10-minute segments to construct SSP maps for each segment across the three studied regions. 
Supplementary Movies 1 and 2 display the SSP maps, normalized within each 10-minute window and across each day for the three regions, respectively. 
Figure 3a presents the average and range of the 10-minute SSP segments over the six days, capturing significant natural and anthropogenic seismic activities as prominent peaks.
These include earthquakes and major transportation movements, such as those involving trains and trucks, which produce energetic seismic waves. 
During this period, we detected all four earthquakes (moment magnitudes, $M_w$: 2.1, 2.7, 2.8, and 2.9) near San Jose, cataloged by the United States Geological Survey.
Additionally, our results reveal fluctuations in urban activities, such as crowd movements around educational institutions. 
At Oak Grove High School, we observe significant increases in SSP coinciding with the school’s bell schedules (Supplementary Fig. 9).
\par
The daily variations in the temporal SSP data reflect the periodicity of human activities. 
On weekdays, the SSP shows distinct peaks during rush hours and lower values at night. 
Additionally, the SSP amplitude is, on average, 1.3 times higher and exhibits 0.8 times less variability on weekdays compared to weekends. 
These findings align with existing literature, confirming that human travel activities are more frequent and regular on weekdays than on weekends~\cite{raux}.
\par
Further validation of our method is provided by the strong correlation observed between SSP and environmental noise levels. 
Urban noise pollution is primarily generated by human activities, especially from vehicular traffic and construction projects~\cite{su12219217,MIR2023104470}, which also produce broadband seismic signals.
A noise level meter was installed near the Seven Trees Branch Library within Region 2. 
Our analysis at this location reveals a significant correlation between urban SSP and environmental noise level (Pearson correlation = 0.75; Spearman correlation = 0.70; Fig. 3b and c).
This strong correlation underscores the validity of our method, demonstrating that heightened SSP values can effectively indicate potential noise pollution hotspots.

\subsection*{Portraying persistent urban features}
\par
The SSP maps (Fig. 4a and Supplementary Movie 2) also portray how the spatiotemporal urban dynamics are influenced by its persistent features, such as land use patterns and demographics.
We find a strong relationship between SSP intensity and land use patterns.
Studies have found that mixed-use lands cultivate urban activities~\cite{tranosmixuse}.
This is validated by our results (Fig. 4a-d): mixed-use zones have the highest level of SSP intensity.
Also, areas allocated for commercial, industrial, and public purposes demonstrate high levels of SSP intensity due to increased urban activities and traffic flows. 
In contrast, areas designated as agricultural or open spaces (A/O) exhibit the lowest SSP intensity, reflecting their reduced urban activity levels. 
Intermediate levels of SSP intensity are noted in residential areas, reflecting a spectrum of urban activities.
Consequently, Region 1 exhibits the lowest SSP intensity, with A/O zones comprising 12.8\% of its land — the highest proportion of A/O land use across the three studied regions (Fig. 4b-d). 
On the other hand, Regions 2 and 3, which have more land for commercial, industrial, and mixed uses, show higher SSP intensity. 
Notably, Region 2, dedicating 10.7\% of its area to commercial use and 5.9\% to mixed-use, exhibits the highest SSP intensity among the studied regions (about two times larger than Region 1). 
These variations highlight how land use patterns influence the urban activities within these zones and, by extension, SSP intensity.
\par
Our results also show that areas with higher residential population density and more visits have higher levels of SSP intensity.
This is verified by the positive correlation between SSP intensity and census block group (CBG)-level population density (Pearson correlation, $r = 0.67$; Spearman correlation, $\rho = 0.73$; Fig. 4e).
A similar positive correlation was found between SSP intensity and average daily traffic ($r = 0.84$; $\rho = 0.77$; Fig. 4f), underscoring the considerable impact of traffic on SSP.
Human activities usually happen in different types of points of interest (POI)~\cite{https://doi.org/10.1111/tgis.12289}.
Our results show that CBGs with higher POI density (number of POIs per squared kilometer) have higher SSP intensity ($r = 0.64$; $\rho = 0.72$; Fig. 4g).
The SSP intensity also positively correlates with visits per squared kilometer ($r = 0.64$; $\rho = 0.60$; Fig. 4h). 
Additionally, we observe that people living in areas that report higher income levels, as derived from CBG-level census data, tend to be exposed to relatively lower SSP intensities, as evidenced by Pearson and Spearman correlations of -0.67 and -0.62, respectively (Fig. 4i).
Overall, our results suggest that after natural seismic sources are removed, urban SSP can be used as a reliable metric for assessing the intensity of urban activities and their spatiotemporal distributions, as well as persistent urban features.

\subsection*{Ubiquitous and general-purpose urban sensing}
\par
The spatial and temporal urban features revealed by our method highlight its capability to achieve ubiquitous and general-purpose urban sensing. 
Our DAS-based urban sensing system offers several advantages over conventional systems, such as those using cameras, vibration sensors, and smartphones. 
Conventional urban sensing systems often suffer from high costs, power and storage requirements, risks of vandalism, and high labor and maintenance expenses. 
While low-power and low-cost sensors offer some solutions~\cite{s150612242,MYDLARZ2017207}, they typically provide only short sensing durations and lack real-time capabilities due to limited network communications. 
Moreover, crowd-sensing approaches using smartphone or vehicular data~\cite{dong2024defining,doi:10.1126/sciadv.1501055}, although cost-effective and at high resolution, suffer from biased sampling and significant privacy concerns.
Our DAS-based system leverages existing telecommunication infrastructure to overcome these challenges. 
By connecting a single optoelectronic instrument (an interrogator) to one end of the fiber and using natural scattering points as seismic sensors spaced every few meters and queried by laser, DAS achieves ultra-dense seismic arrays at a cost of only a few dollars per meter~\cite{10.1785/0220190112}. 
This system ensures privacy by avoiding the collection of identifiable information and enables real-time and continuous sensing through regular network communications.
Thanks to its ultra-dense array property, our system can effectively detect and locate urban activity-related seismic sources -- a task challenging for current seismic sensing networks, as high-frequency seismic waves are subject to scattering and attenuation over short distances~\cite{https://doi.org/10.1002/2015GL063558}.
\par
Furthermore, the existing telecommunication infrastructure supports our DAS implementation, enabling ubiquitous urban sensing. 
By using seismic interferometry and beamforming, each DAS array can cover an area extending hundreds to thousands of meters around it to remotely map seismic sources.
This capability overcomes the proximity sensing limitations of previous urban science applications using DAS. 
In the United States, extensive fiber optic networks -- buried underground or suspended from poles -- connect homes, businesses, and data centers in most cities~\cite{fcc2023}.
A recent study demonstrates that integrating existing Internet fiber-optic cables as seismic sensors can significantly increase the monitored area of US metropolitan statistical areas for low-amplitude ground-motion events (i.e., moment magnitude $>0.5$), expanding coverage from an average of 1\% to 12\%~\cite{10.1785/0220240049}. 
This study uses 1.5 km as the detection threshold of sensors in fiber-optic cables for events of magnitude 0.5 or greater, which aligns with our sensing range for mapping seismic sources like trucks. 
The sensing coverage estimation derived in~\cite{10.1785/0220240049} also applies to our approach, thus enabling our ubiquitous urban sensing.
\par
Urban seismic signals, originating from sources like ground transportation, industrial activities, cultural events, and natural phenomena like earthquakes and landslides, provide valuable spatial and temporal urban data.
Urban seismic signals and SSP maps are instrumental in characterizing human activities and monitoring urban environments.
We have demonstrated correlations between SSP intensity and various urban features, establishing a proxy for assessing urban environments and society.
They could aid in optimizing urban layouts, improving traffic management, and reducing vibration and noise pollution, which can potentially enhance urban livability and reduce adverse health impacts on residents~\cite{RanpiseTandel+2022+48+66,Sun_2016}. 
Our study also has the potential to enhance urban security by providing critical insights into natural and anthropogenic hazards, including earthquakes~\cite{10.1785/0320230018}, infrastructure failures~\cite{liu2023turning}, and man-made blasts~\cite{doi:10.1190/segam2017-17745041.1}.
Additionally, both natural and anthropogenic seismic sources have proven effective in probing subsurface structures for passive seismic surveys~\cite{spica2020urban, fang2020urban}. 
By mapping these sources, we can better characterize subsurface structures and effectively assess seismic risks in urban areas, thereby improving disaster preparedness and supporting the development of resilient civil infrastructure~\cite{seismic1,Xu2022}.
\par
By demonstrating the technological feasibility and cost-effectiveness of a ubiquitous, general-purpose urban sensing platform, this study empowers the urban science community with a tool to enhance the understanding and modeling of complex urban environments and society.

\section*{Methods}
\subsection*{Data description and pre-processing}
\par
This study uses DAS data collected over six days, from September 20th to 25th, 2023, in San Jose, California. 
Data acquisition was conducted using a QuantX interrogator provided by Luna - OptaSense~\cite{quantx}, which functioned at a sampling frequency of 200 Hz and was configured with a gauge length of 10 meters. 
The interrogated telecommunication fiber-optic cable spans approximately 50 kilometers in length. 
This setup facilitated 50,000 distributed dynamic-strain sensors along the whole cable, with each sensor spaced 1 meter apart. 
For seismic source mapping, we identified three specific regions, as shown in Fig 1a. 
These regions were selected based on the layout of the telecommunication cables, which form a loop within each area, thereby providing a geometric constraint for beamforming. 
Specifically, the lengths of the telecommunication cables in regions 1, 2, and 3 measure 9.5 km, 4 km, and 5.6 km, respectively, which all fall within the operational sensing range of the DAS system.
\par
The preprocessing of DAS signals involves two primary steps: bandpass filtering and excluding outliers. 
After signal detrending, a bandpass filter (1 to 20 Hz) is applied. 
This band is selected to include urban seismic signals from sources of interest above 1 Hz while filtering out low-frequency quasi-static signals, high-frequency noise, and near-field signals. 
Subsequently, the data is segmented into 10-minute windows for analyzing the temporal variations in SSP. 
To mitigate the effects of large amplitude outliers, such as those caused by direct impacts on the fiber, signals in each window that exceed the 99th percentile in amplitude are replaced with the median value obtained from a spatial window of 50 DAS channels. This window is centered on the channel being replaced at the corresponding time.
Furthermore, to reduce computational costs while preserving sufficient spatial resolution, DAS signals are subsampled by a factor of 10 in the spatial domain.

\subsection*{Seismic source power estimation with DAS} 
\par
We employ the beamforming technique~\cite{10.1155/2012/292695,Dougherty2002} to estimate the power of urban seismic sources at different locations by shifting and stacking the DAS recorded seismic traces according to a wave propagation model. 
The beamforming approach aims to estimate the coherent wave energy traversing the DAS array and to characterize its propagation attributes~\cite{krim1996two}.
Our analysis focuses on analyzing the propagation of surface waves to locate urban seismic sources.
The beamforming is performed in the frequency domain 
because of the dispersion characteristics of surface waves, where the wave velocity varies with frequency~\cite{eage:/content/journals/10.3997/1873-0604.2004015}.
Our beamforming implementation is based on the software package Acoular~\cite{SARRADJ201750}
\par
Initially, we consider the analysis of a single seismic source located at $\mathbf{x}_s$ using a DAS array with $N$ channels.
The seismic signal recorded by the $r$-th DAS channel at position $\mathbf{x}_r$ and frequency $\omega$ is:
\begin{equation}
p(\omega;\mathbf{x}_r) = a(\omega;\mathbf{x}_r,\mathbf{x}_s) s(\omega),
\end{equation}
where $s(\omega)$ denotes the seismic amplitude characterized at the source location at frequency $\omega$, and ${a}(\omega,\mathbf{x}_r,\mathbf{x}_s)$ is the transfer function that accounts for signal attenuation and phase delays at frequency $\omega$. 
Assuming a monopole source with a frequency-independent attenuation, the transfer function is defined as:
\begin{equation}
a(\omega;\mathbf{x}_r,\mathbf{x}_s) = \frac{1}{\sqrt{r_{s,r}}}\exp\left(-\frac{\omega{r}_{s,r}}{2Qv_{\omega}}\right) \exp\left(-\frac{i\omega {r}_{s,r}}{v_{\omega}}\right),
\label{eq:transfer}
\end{equation}
where ${r}_{s,r} = |\mathbf{x}_s - \mathbf{x}_r|$ denotes the Euclidean distance between the source and the DAS channel at $\mathbf{x}_r$, $v_{\omega}$ is the wave propagation velocity at frequency $\omega$, and $Q$ is the attenuation quality factor. Phase distortion due to the attenuation is not considered here.
The vector of seismic signals at the DAS array at frequency $\omega$ due to a source at $\mathbf{x}_s$ is given by $\mathbf{p}(\omega) = \mathbf{a}(\omega;\mathbf{x}_s) s(\omega)$, where vector $\mathbf{a}(\omega;\mathbf{x}_s)$ includes all individual transfer functions and compensates for the time delays and attenuation of the seismic wave traveling from the source to the $N$ DAS channels. 
\par
The estimated seismic power at a position $\mathbf{x}_t$ and frequency $\omega$ is estimated using the real-value auto-power spectrum:
\begin{equation}
S(\omega,\mathbf{x}_t) = \mathbf{h}^H(\omega;\mathbf{x}_t) \mathbb{E}\left[\mathbf{p}(\omega)\mathbf{p}^H(\omega)\right] \mathbf{h}(\omega;\mathbf{x}_t),
\label{eq:power}
\end{equation}
where the superscript $H$ denotes the Hermitian transpose, and $\mathbb{E}[\cdot]$ is the expectation operator. 
The cross-spectral matrix $\mathbb{E}\left[\mathbf{p}(\omega)\mathbf{p}^H(\omega)\right]$ has dimensions $N \times N$, with each element at the $i$-th row and $j$-th column corresponding to the cross-spectrum density function of the signals between the $i$-th and $j$-th DAS channels at frequency $\omega$. 
The cross-spectrum function is the Fourier transform of the cross-covariance function of the time-domain signals between these channels. 
The steering vector $\mathbf{h}(\omega,\mathbf{x}_t)$ is to calculate the weighted sum of the DAS signals using complex-valued weight factors at the location $\mathbf{x}_t$ and frequency $\omega$.
Applying this steering vector maximizes output power when the assumed and actual source positions match: $S(\mathbf{x}_t = \mathbf{x}_s) > S(\mathbf{x}_t \neq \mathbf{x}_s)$. 
The calculated output power also should approximate the source power, i.e., $S(\omega;\mathbf{x}_t = \mathbf{x}_s) \approx \mathbb{E}[s(\omega)s^H(\omega)]$. 
Various steering vector formulations exist; we adopt the formulation as suggested in~\cite{MA2020115064,KARIMI2023110454}:
\begin{equation}
\mathbf{h}(\omega;\mathbf{x}_t) = \frac{1}{\sqrt{N}} \frac{\mathbf{a}(\omega,\mathbf{x}_t)}{\sqrt{\mathbf{a}^H(\omega;\mathbf{x}_t) \mathbf{a}(\omega;\mathbf{x}_t)}}.
\end{equation}
\par
In our analysis, we make two additional assumptions. First, when performing beamforming, the directionality effect of DAS is not considered. 
DAS records the combined horizontal components of the wavefield, influenced by incident angle and cable orientation. 
Characterizing directional sensitivity is challenging due to unknown source polarization. 
Moreover, seismic signals from multiple sources are considered to be uncorrelated, allowing $S(\omega;\mathbf{x}_t)$ to sum their contributions. 
This algorithm remains valid even in scenarios involving multiple sources.
\par
Prior to beamforming, it is necessary to estimate two key parameters for each studied region: wave propagation velocity and the attenuation quality factor. 

\subsection*{Surface wave velocity estimation}
\par
To estimate surface wave velocities, we construct Virtual Shot Gathers (VSGs) from vehicle-induced surface waves recorded by DAS.
As vehicles travel along a roadside DAS array, they produce two types of signals: the quasi-static deformation due to the vehicles' weight and the vehicle-induced surface waves, predominantly Rayleigh waves. 
To utilize these surface waves, we first separate the quasi-static deformation (below 1 Hz) and surface waves (1-20 Hz) using low-pass and bandpass filtering of the DAS data, respectively. 
The quasi-static deformation signals are used to track the locations of moving vehicles on the DAS record through a specialized Kalman filter~\cite{10.1145/3596262}. 
This algorithm calculates the arrival times of vehicles at each DAS channel using a prominence-based peak detection method and recursively determines the posterior probability of vehicle arrival times across space (in the direction of vehicle motion) by integrating spatial-dependent vehicle detection results from multiple DAS channels.
Vehicle tracking results are obtained by converting the estimated arrival times into vehicle locations. 
\par
The tracked vehicle trajectories on the DAS records further enable us to select corresponding surface-wave windows (Supplementary Fig. 1a and b). 
To avoid interference from nearby vehicle-induced wavefields, we select isolated vehicles with a minimum separation of 25 seconds between other vehicles.
We construct VSGs by performing cross-correlations of the pivot trace with other traces within these surface-wave windows. 
Vehicles passing by generate both forward- and backward-propagating waves on either side of the vehicle's trajectory (Supplementary Fig 1c). 
We distinctly handle the positive and negative offset sections of wave interferometry~\cite{https://doi.org/10.1029/2023JB028033}. 
For the backward-propagating waves, i.e., waves traveling in the direction opposite to the vehicle movement, we define the cross-correlation function $C\left(\mathbf{x}_s, \mathbf{x}_r, \tau \right)$ as
\begin{equation}
C\left(\mathbf{x}_s, \mathbf{x}_r, \tau \right) = \left\{\begin{array}{l}
\int_{t_s+\epsilon}^{t_s+\epsilon+\Delta t} u\left(t+\tau; \mathbf{x}_r\right) \cdot u\left(t; \mathbf{x}_s\right) d t, \quad \mathbf{x}_r < \mathbf{x}_s \\
\\
\int_{t_r+\epsilon}^{t_r+\epsilon+\Delta t} u\left(t-\tau; \mathbf{x}_r\right) \cdot u\left(t; \mathbf{x}_s\right) d t, \quad \mathbf{x}_r \geq \mathbf{x}_s,
\end{array}\right.
\end{equation}
where $u\left(t; \mathbf{x}\right)$ is the recorded DAS trace at time $t$ and location $\mathbf{x}$, $C\left(\mathbf{x}_s, \mathbf{x}_r, \tau \right)$ denotes the inter-channel correlation between recorded DAS strain wavefields at two channel pairs $\mathbf{x}_s$ and $\mathbf{x}_r$ with $\tau$ denoting the time lag. 
Here, $C\left(\mathbf{x}_s, \mathbf{x}_r, \tau \right)$ approximates the wavefield as if we place a source at $\mathbf{x}_s$ and a receiver at $\mathbf{x}_r$. 
For the negative-offset section in VSGs, channels traversed by the vehicle ($\mathbf{x}_r < \mathbf{x}_s$) are cross-correlated with the pivot trace $u(t, \mathbf{x}_s)$ within the time window $[t_s+\epsilon, t_s+\epsilon+\Delta t]$.
Here, $ t_s $ denotes the vehicle's arrival time at the virtual source location $ \mathbf{x}_s $, $\Delta t$ denotes the selected time window length for cross-correlation, and $\epsilon$ is a time lag introduced to avoid near-field effects. 
For the positive-offset section of VSGs, the cross-correlation is performed in the time window of $[t_r+\epsilon, t_r+\epsilon+\Delta t]$, where $t_r$ represents the arrival time of the traveling vehicle at virtual receiver location $\mathbf{x}_r$. 
In the case of forward-propagating waves, we employ the time window $[t_r-\epsilon-\Delta t, t_r-\epsilon]$ for the negative-offset section ($\mathbf{x}_r < \mathbf{x}_s$), and $[t_s-\epsilon-\Delta t, t_s-\epsilon]$ for the positive-offset segment ($\mathbf{x}_r \geq \mathbf{x}_s$). Then, we stack VSGs from many individual vehicles at the same location to form the final VSGs to enhance the signal-to-noise ratio. 
\par
The approach we used here to build the VSGs is more computationally efficient than the conventional ambient noise approach, as only windowed data is required for the cross-correlation. 
Moreover, this method yields more robust VSGs construction without relying on the assumption of uniform source distribution in the ambient noise interferometry~\cite{yuan2020near}. 
\par
We perform the aforementioned algorithm to construct VSGs along the fiber cable across all three regions of interest.
We obtain VSGs at 2, 5, and 8 locations by stacking 485, 2,520, and 3,492 vehicles along the fiber in Region 1, 2, and 3, respectively.
For all three regions, each VSG utilized 300 DAS channels with channel spacing of 1 m, corresponding to an offset range of -150 m to 150 m.
The variation in the number of vehicles isolated for VSGs is due to differences in fiber layouts and traffic volumes across the studied regions. 
In particular, Region 1 experiences less traffic compared to the other two regions.
Supplementary Fig. 2a-c show the stacked VSGs in the three regions, and Supplementary Fig. 2d-f show the phase velocity dispersion images via the phase-shift method~\cite{doi:10.1190/1.1444590}. 
The anti-causal component of the VSGs (negative time lag) typically originates from vehicles in distant opposing traffic lanes~\cite{https://doi.org/10.1029/2023JB028033} and is excluded from the dispersion analysis.

\par
For our beamforming algorithm, we use the fundamental mode dispersion curve in each region, assuming a uniform velocity model within each region. 
These dispersion curves are estimated by averaging the results across all VSG locations within each region, resulting in a single dispersion curve per region. 
The fundamental mode dispersion curves for surface waves derived from the VSGs are shown in Supplementary Fig. 2g–i.
The study area exhibits geological diversity, as illustrated in the Supplementary Fig. 3.
In the computed frequency range (1.5 Hz to 8 Hz), Region 1 exhibits an S-wave velocity structure ranging from around 650 m/s to 400 m/s, whereas Region 2 displays a broader velocity profile, spanning from 750 m/s to 320 m/s. Region 3 has lower velocity structures compared to the other two regions.
Based on these dispersion curves, we do not observe significant variations in surface wave velocity across different locations within each region (gray lines), supporting the adequacy of using a single surface wave velocity profile for our current approach. 
Also, we focus on low frequencies (1.5 Hz to 8 Hz), which propagate through the less variable parts of the subsurface compared to the top few meters.

\subsection*{Attenuation quality factor estimation}
\par
The attenuation quality factor, $Q$, is another essential parameter for computing the transfer function in Eq.~\ref{eq:transfer} before estimating the seismic source power. 
We assume a constant attenuation quality factor~\cite{https://doi.org/10.1029/JB084iB09p04737} across the subsurface in each region. 
The logarithm of the spectral ratio between two seismic signals at different locations can be expressed as:
\begin{equation}
\ln\left(\frac{A(\omega;\mathbf{x}_i)}{A(\omega;\mathbf{x}_j)}\right) = -\frac{\omega {r}_{i,j}}{2Qv_{\omega}} + C,
\label{eq:logspectral}
\end{equation}
where $A(\omega;\mathbf{x}_i)$ and $A(\omega;\mathbf{x}_j)$ represent the energy of seismic waves at frequency $\omega$ at locations $\mathbf{x}_i$ and $\mathbf{x}_j$, respectively.
The constant $C$ is an intercept term, and ${r}_{i,j} = |\mathbf{x}_i - \mathbf{x}_j|$ denotes the Euclidean distance between the two locations.
\par
In our analysis, seismic signals induced by trucks are utilized to estimate the attenuation factor because trucks are prevalent and strong seismic sources in all three studied regions.
Also, the truck-induced seismic signals exhibit a broader peak in the dominant frequency response compared to other urban seismic sources, such as trains or construction activities (Supplementary Fig. 4).
The first step in estimating the attenuation factor involves transforming the truck-induced seismic signals to the frequency domain using a Fourier transform for each DAS signal.
Based on Eq.~\ref{eq:logspectral}, $Q$ is then estimated by solving the following least square problem:
\begin{equation}
\hat{Q}, \hat{C} = \arg\min_{C, Q} \sum_{\omega} \sum_r \left[\ln\left(\frac{A_0(\omega)}{A_r(\omega)}\right) + \frac{\omega r}{2Qv_{\omega}} - C\right]^2,
\label{eq:attenuation}
\end{equation}
where $A_0(\omega)$ is the reference power spectral density (PSD) of DAS signals at frequency $\omega$. 
This reference energy is computed as the average PSD from DAS channels located at a distance of 50 to 90 meters from the truck source to avoid signal clipping near the source that could underestimate the energy.
$A_r(\omega)$ is the average PSD of signals from DAS channels within a distance of $r-d/2$ to $r+d/2$ from the reference signals' center, with $d$ set at 10 meters.
This calculation also aids in assessing the sensing coverage of our DAS system for various urban seismic sources (Supplementary Fig. 5).
Supplementary Fig. 6a-c show the attenuation of truck-induced seismic power in the 2.8-3.4 Hz range across the three studied regions. 
The plots in Supplementary Fig. 6d-f show the logarithmic spectral ratios at varying distances from the truck source, where the slope of the blue dashed line is proportional to $Q^{-1}$.
The estimated attenuation quality factors for regions 1, 2, and 3 are 6, 8, and 14, respectively. 
\par
It is important to note that DAS signal amplitude is influenced not only by subsurface structures but also by the ground coupling of the fiber-optic cable and the directional sensitivity of the wavefield. 
The lack of coupling data across the telecommunication network limits our ability to isolate attenuation solely attributed to subsurface structures.
Furthermore, accounting for the directional sensitivity of the wavefield requires detailed information on source polarization~\cite{martin2021introduction}, which is often challenging to obtain for passive sources. 
Therefore, the attenuation determined reflects a composite effect of subsurface characteristics, cable coupling, and directionality. 
In this context, the derived $Q$ is an approximation for these multifaceted influences.

\subsection*{Seismic source mapping}
\par
To construct spatial maps of SSP across the three studied regions (Fig. 4a), we have partitioned each area into a grid layout. 
Each grid cell within these grids measures 50 meters by 50 meters, a size chosen to balance the granularity of the spatial map. 
This ensures that each cell adequately represents local variations while maintaining manageable data volumes for processing.
The power spectrum of the seismic source in the frequency range of 1.5 to 8 Hz for each grid cell is estimated using Eq.~\ref{eq:power}. 
This specific frequency range is selected to effectively capture the primary frequencies associated with urban seismic activities while excluding high-frequency near-field noise. 
This choice is further supported by observations that many common urban activities display dominant frequencies within this spectrum (Supplementary Fig. 4).
To minimize the influence of near-field sources, such as vehicles traversing directly above the fiber, the seismic source power for each grid cell is estimated using DAS channels located more than 100 meters away from that grid cell.
Seismic source mapping can be conducted for a specific frequency of interest or across the entire frequency band by aggregating the maps of each individual frequency. Supplementary Fig. 7 visualizes the seismic source intensity (seismic source power per unit of area) of each census block, with the hatch pattern indicating the land use zoning categories. 
\par
The accuracy and effectiveness of our seismic source mapping are verified through a series of validations. 
These tests compare the estimated source locations with known coordinates of various urban activities. 
The validation scenarios include tracking moving trains (e.g., Supplementary Fig. 8 and Fig. 2a, d), locating trucks (e.g., Fig. 2b and e), identifying a construction site (Fig. 2c and f), and monitoring activities around a school district (Supplementary Fig. 9).
We observe high SSP values aligning with the shape of the fiber array, primarily because the fiber routes follow major roads, such as arterial streets and highways, in each region. These roads experience higher traffic volumes, generating more traffic-induced seismic power compared to smaller roads or areas farther from the fiber array. 
Consequently, the observed high-power regions largely reflect the alignment of seismic source hotspots along these major roads.
Furthermore, SSP maps are calculated for each 10-minute segment of the DAS data. 
These maps illustrate the daily rhythms and spatial variations of urban dynamics (Supplementary Movies 1 and 2).
\par
There are several limitations and future directions of our methodology that should be noted.
First, accurate seismic source localization using the DAS system requires coverage from telecommunication optical fibers across sufficient back azimuths for effective beamforming. 
A better result is achieved with telecommunication optical fibers encircling the studied regions, creating a geometric constraint essential for effective beamforming. 
For instance, it would be challenging to determine the source locations using only a straight fiber cable.
Also, our analysis reveals that the sensing range of DAS data varies with the energy and frequency of different sources (Supplementary Fig. 5).
It is difficult to accurately detect weak and high-frequency seismic sources distant from the optical fiber network. 
Nonetheless, the extensive urban telecommunication network offers the potential for multi-back azimuth coverage and dense sensing infrastructure.
Future research could explore the optimal configuration of optical fibers to maximize the accuracy of seismic source mapping and quantify the sensing coverage for different sources. 
\par
Furthermore, although our methodology utilizes distant channels to reduce the influence of near-field effects, the attenuation of far-field sources can still bias the spatial distribution of energy toward near-field activity. Future work should aim to address these limitations by developing techniques to differentiate between near-field and far-field contributions more effectively.
While regional heterogeneity of subsurface properties, including the surface wave velocity and attenuation, are taken into account, improving the spatial resolution of these properties could enhance the accuracy of seismic source mapping. 
Future work should also consider the complexities associated with existing dark fibers, including variations in coupling conditions and conduit materials. Addressing these challenges could involve developing efficient calibration methods. For example, controlled driving tests using vehicles with known weights, combined with precise information about the fiber’s location and depth, could provide a practical and scalable solution.
Looking ahead, as our urban sensing approach leveraging existing fiber-optic networks gains traction, the expanded spatial coverage and increased density of these networks will provide richer datasets and finer spatial resolution, enabling the development of more accurate and detailed three-dimensional surface wave velocity models.
It is also worthwhile to explore alternative beamforming algorithms, which could potentially improve spatial resolution and enhance side-lobe suppression~\cite{10.1155/2012/292695}.
Given the challenges associated with the unknown characterization of urban sources, we do not include the effects of fiber and source directionality in our model. Future studies, particularly those involving controlled experiments, could provide valuable insights into the impact of source characteristics and DAS directional sensitivities on source mapping results.
The recorded seismic signals demonstrate distinct patterns for various types of urban events (Supplementary Fig. 4).
This suggests that pattern recognition and machine learning algorithms could be applied to analyze massive amounts of DAS data, enabling efficient and automated monitoring of diverse urban activities.

\section*{Data availability}
Supplementary Information is available for this paper.
The DAS recordings used for validating the seismic source mapping are available at \url{https://doi.org/10.5281/zenodo.12725788}.
Data on land use, construction sites, population density, average daily traffic, and median income are publicly accessible through the San Jose CA Open Data Portal (\url{https://data.sanjoseca.gov/}). 
The POI density and visit data are accessible and can be requested for research purposes at \url{https://www.safegraph.com/}.

\section*{Code availability}
The code for processing the DAS data and performing seismic source mapping is available at~\cite{urban_das}.
The code for estimating surface wave velocity is adopted from~\cite{https://doi.org/10.1029/2023JB028033} and is available at~\cite{das_veh}.


\section*{Acknowledgements}
LUNA-Optasense contributed to the recording of DAS data in San Jose by loaning the instrument and providing field support. 
In particular, Andres Chavarria facilitated the instrument loaning, and Victor Yartsev helped in the field operations.
We also thank Dubai Future Foundation, UnipolTech, Consiglio per la Ricerca in Agricoltura e l’Analisi dell’Economia Agraria, Volkswagen Group America, FAE Technology, Samoo Architects \& Engineers, Shell, GoAigua, ENEL Foundation, Kyoto University, Weizmann Institute of Science, Universidad Autónoma de Occidente, Instituto Politecnico Nacional, Imperial College London, Universitá di Pisa, KTH Royal Institute of Technology, AMS Institute and all the members of the MIT Senseable City Lab Consortium for supporting this research. 
We appreciate the support from the Stanford Doerr School of Sustainability and the Center for Computation in handling the large datasets generated by DAS systems. 
We thank Robert Clapp, Siyuan Yuan, Thomas Cullison, Jatin Aggarwal, Doyun Hwang, and Seunghoo Kim from Stanford University for helping with the data collection, as well as the City of San Jose -- particularly Darren Thai and Ho Nguyen -- for crucial help with the experiment. 
We thank Chada Elalami, Isabel Waitz, and Sabrina Tian from the Senseable City Laboratory at MIT for their help in enhancing figures and creating the featured image.

\section*{Author contributions}
J.L., B.B., and P.S. defined the problem.
J.L. and B.B. designed the research. 
J.L. collected the DAS data, supervised by H.N. and B.B.
J.L. performed the analysis, supervised by B.B., P.S., and C.R.
J.L. and H.L. designed the attenuation quality factor estimation method.
J.L. implemented the processing code. 
J.L., H.L., and P.S. prepared the manuscript. 
B.B., P.S., H.N., and C.R. jointly revised the paper and supervised the research.
All authors reviewed the manuscript.

\section*{Competing interests}
Dubai Future Foundation, UnipolTech, Consiglio per la Ricerca in Agricoltura e l’Analisi dell’Economia Agraria, Volkswagen Group America, FAE Technology, Samoo Architects \& Engineers, Shell, GoAigua, ENEL Foundation, Kyoto University, Weizmann Institute of Science, Universidad Autónoma de Occidente, Instituto Politecnico Nacional, Imperial College London, Universitá di Pisa, KTH Royal Institute of Technology, AMS Institute and all the members of the MIT Senseable City Lab Consortium do not receive any direct financial or economic advantage from the publication of this study and had no role in study design, data collection and analysis, decision to publish or preparation of the manuscript. 
Furthermore, neither LUNA-OptaSense nor its employees, including Andres Chavarria and Victor Yartsev, received any direct financial or economic advantage from the publication of this study.
The authors declare no competing interests.

\FloatBarrier
\section*{Figures}
 \begin{figure}[!htb]
    \centering
    \includegraphics[width=1\textwidth]{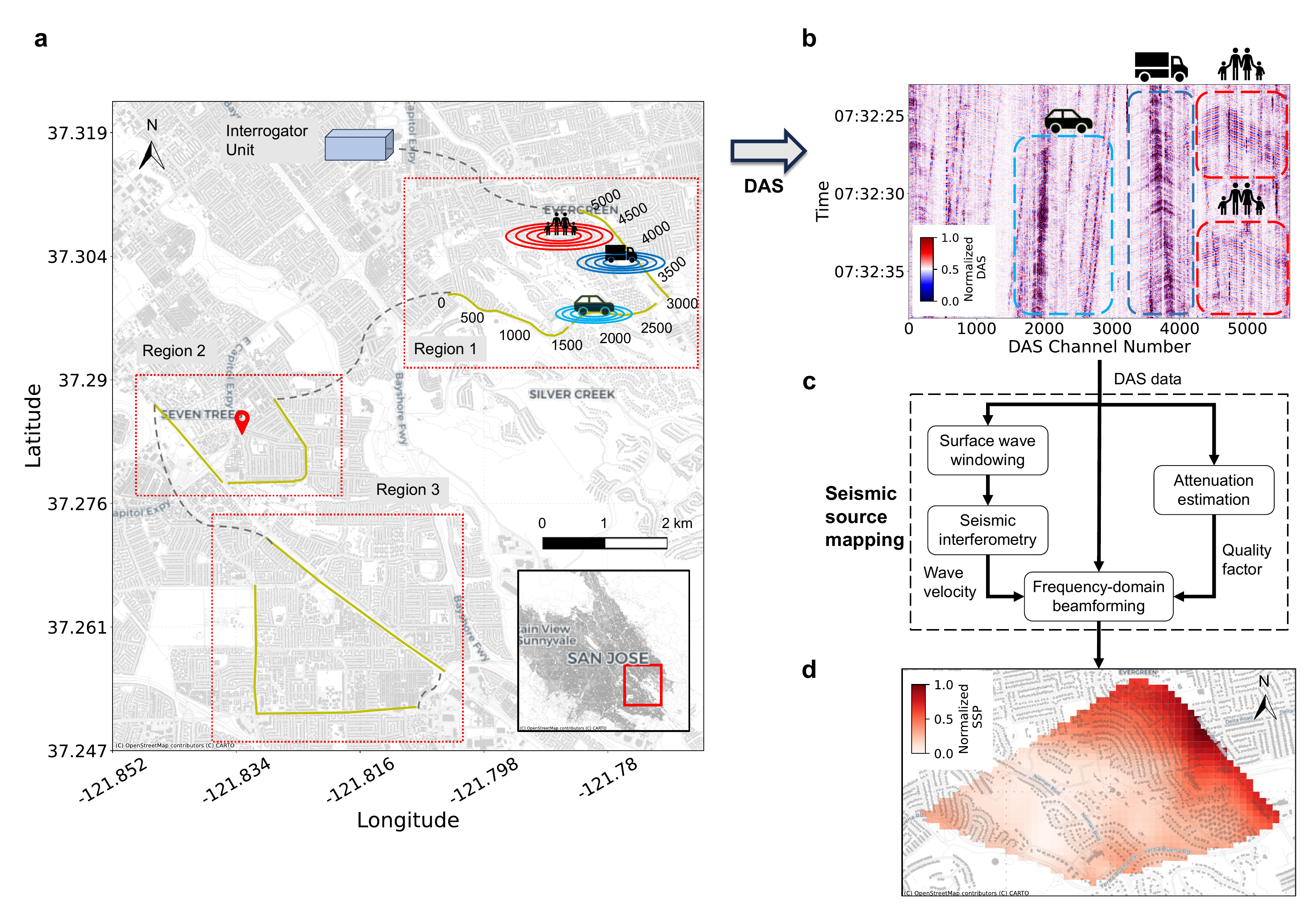}
    \caption{{\bf Seismic source mapping using existing telecommunication fiber-optic cables as seismic sensing arrays using distributed acoustic sensing (DAS).} 
{\bf a}, Map of the DAS setup in San Jose, California, showing an interrogator unit connected to existing telecommunication fiber-optic cables. 
Dotted boxes outline the three regions studied, with yellow lines indicating the fibers utilized for DAS. 
Gray dashed lines suggest connectivity of these fibers rather than the actual fiber routes. 
A noise level meter was placed at the Seven Trees Branch Library as marked by the red pin. 
The comparison between the estimated seismic source power and environmental noise levels is shown in Fig. 3b and c.
Region 1 is highlighted to demonstrate our method, with DAS channel annotated spacing at 1-meter intervals.
Our method processes the acquired data, as shown in {\bf b}, to estimate a heat map of SSP, as shown in {\bf c}. 
{\bf b}, A snapshot of DAS data displaying spatial and temporal seismic activities along the fiber array. The segments from channels 1750 to 2750 and 3250 to 4250 register high-frequency seismic waves due to vehicle movements, while the span from channels 4000 to 5500 records lower-frequency waves from residential areas. 
{\bf c}, A block diagram illustrating our method for seismic source mapping.
{\bf d}, A heat map calculated from the processed DAS data shows the SSP across the area.
It is noted that lower-frequency seismic waves have a farther propagation range than high-frequency waves, and both urban activities are identified in the SSP map shown in {\bf d}.}
    \label{fig:system}
\end{figure}

\begin{figure}[!htb]
    \centering
    \includegraphics[width=1\textwidth]{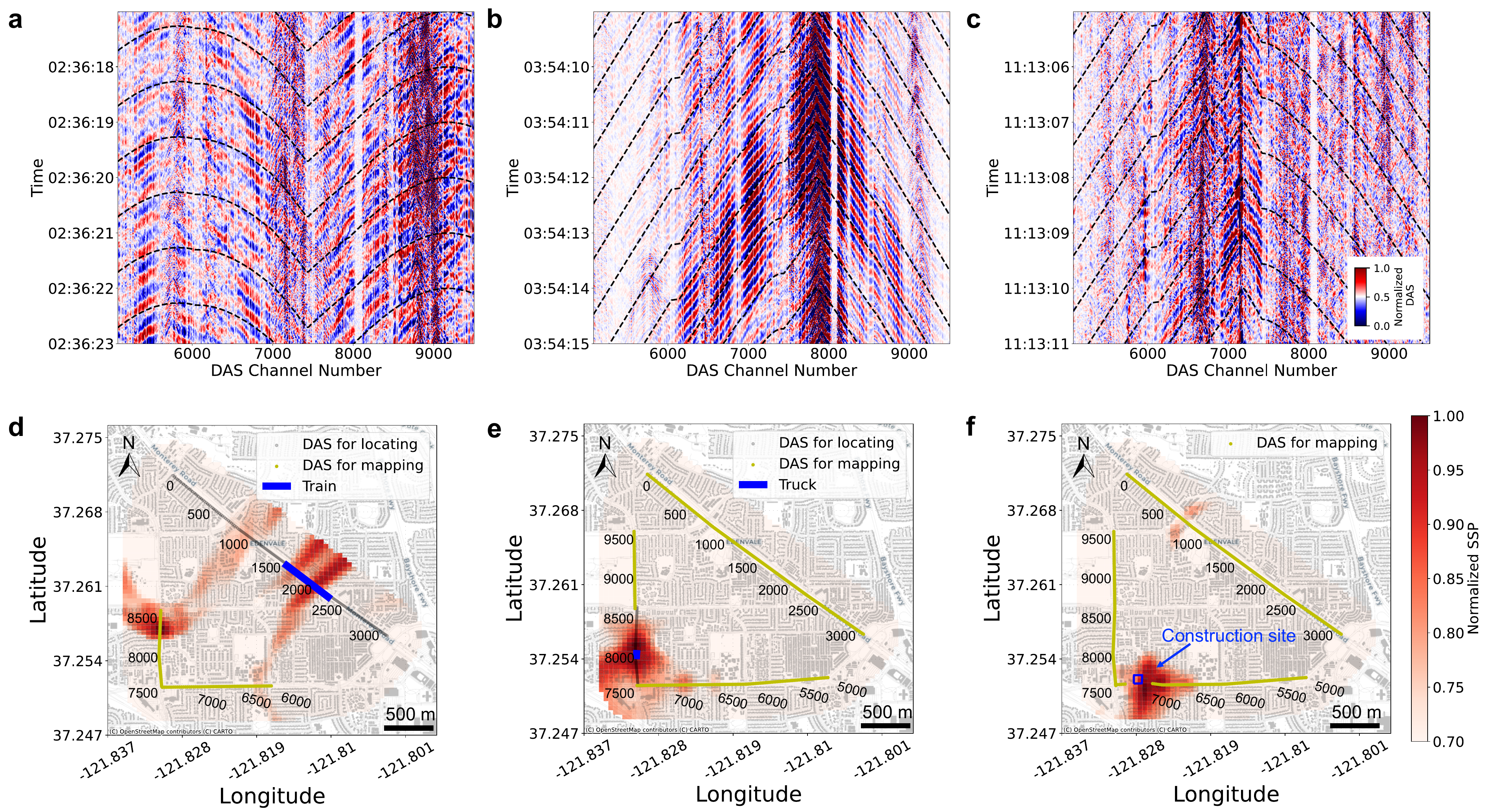}
    \caption{{\bf Validations of seismic source mapping using various seismic activities with known coordinates and timings.}
Using fibers aligned with roadways and railways, we obtain positions and timings of seismic sources, such as trains and trucks passing, as ground truth. 
We use DAS data from remote fibers to detect and locate these seismic sources through beamforming.
{\bf a}, Normalized seismic records (proportional to strain) from the fiber segments capture a train passing event.
The black dashed lines represent the estimated wave arrival times, calculated every second from the located source.
{\bf d}, SSP map during the train passing, with the train's position (solid blue line) detected from the quasi-static signal captured by the DAS array adjacent to the railway track (Channels 0 to 3,500). 
Gray dots show DAS channels for locating the train with labeled channel numbers. 
Yellow dots highlight the active DAS channels used for seismic source mapping. 
The estimated seismic source hotspot aligns with the actual train location across channels 1,500 to 2,250. 
Another seismic hotspot in the north is due to roadway traffic. 
{\bf b} and {\bf e} extend the analysis to a truck movement along the fiber pathway, demonstrating our ability to detect seismic waves generated by the truck from up to 1.5 kilometers away (Supplementary Fig. 5b). 
{\bf c} and {\bf f} apply the same methodology to a construction site, indicated by a blue box, where seismic waves generated by construction activities are detectable by our system up to 500 meters (Supplementary Fig. 5c). 
The estimated arrival times of these seismic events correspond closely with the actual waveform, and the mapped seismic source hotspots align with the real locations of the seismic events, thereby validating the accuracy of our method in mapping both moving and stationary seismic sources.}
    \label{fig:DOAs}
\end{figure}

\begin{figure}[!htb]
    \centering
    \includegraphics[width=1\textwidth]{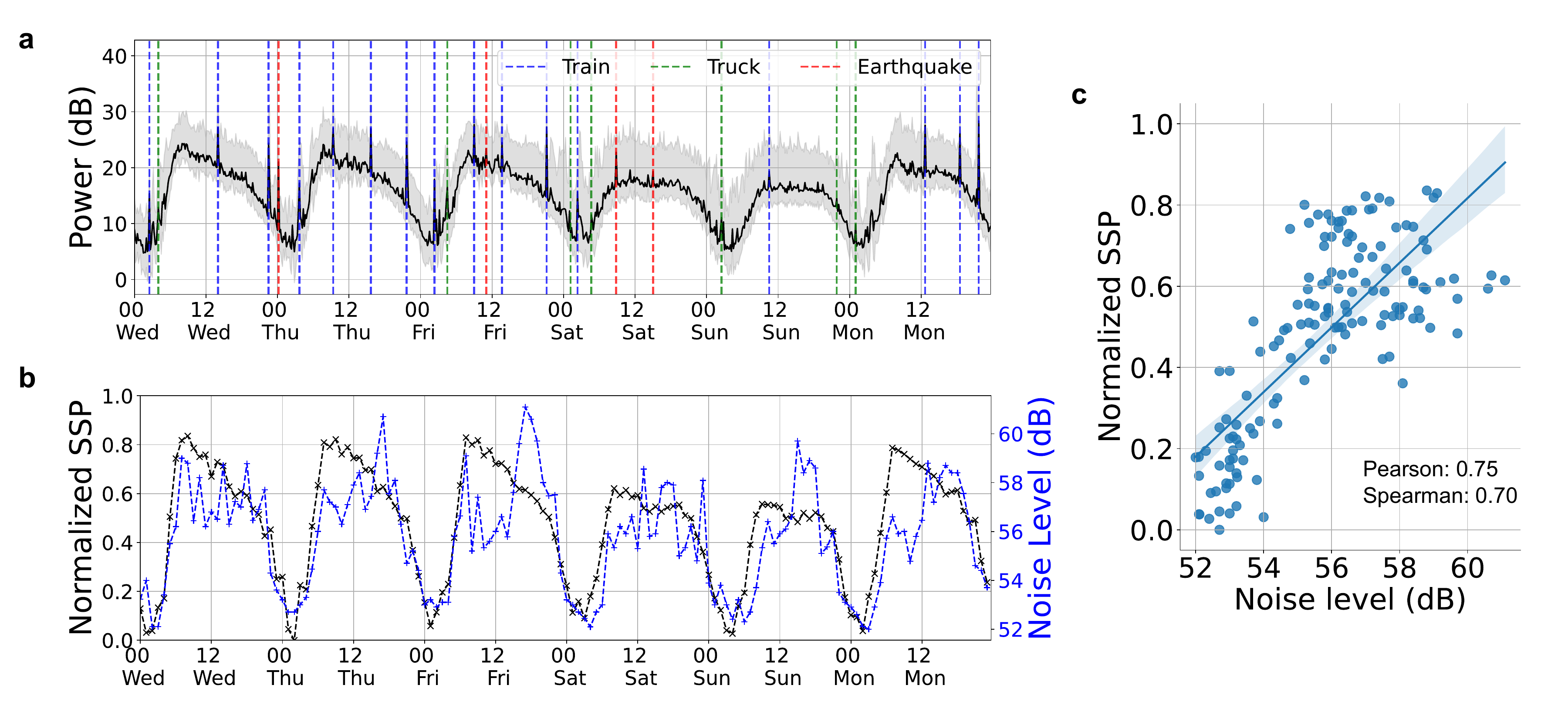}
    \caption{{\bf Temporal variations of urban SSP and its correlation with environmental noise level measurements.} 
{\bf a}, Time series plot displaying the seismic source power over six days (from September 20th to 25th, 2023) in Region 3, showing the range of power values as shaded areas. The temporal resolution is 10 minutes. Vertical dashed lines pinpoint energetic regional seismic events, including train and truck passing and earthquakes. The periodicity in the data reflects the influence of daily human activities on seismic source power, with notable peaks during rush hours and declines over weekends. 
{\bf b}, Time series plot illustrating the variations in hourly seismic source power, depicted by the black line, and environmental noise levels, represented by the blue line, over a six-day period in Region 2. 
The ambient noise level meter was placed near the Seven Trees Branch Library (as indicated by a red pin in Fig. 1a). The rhythmic fluctuations in seismic activity correspond closely with environmental noise patterns, mirroring the daily rhythms of environmental noise at this location. 
{\bf c}, A scatter plot demonstrating the correlation between normalized SSP and noise level, with Pearson and Spearman coefficients highlighting a significant positive relationship.}
    \label{fig:temporal}
\end{figure}

\begin{figure}[!htb]
    \centering
    \includegraphics[width=1\textwidth]{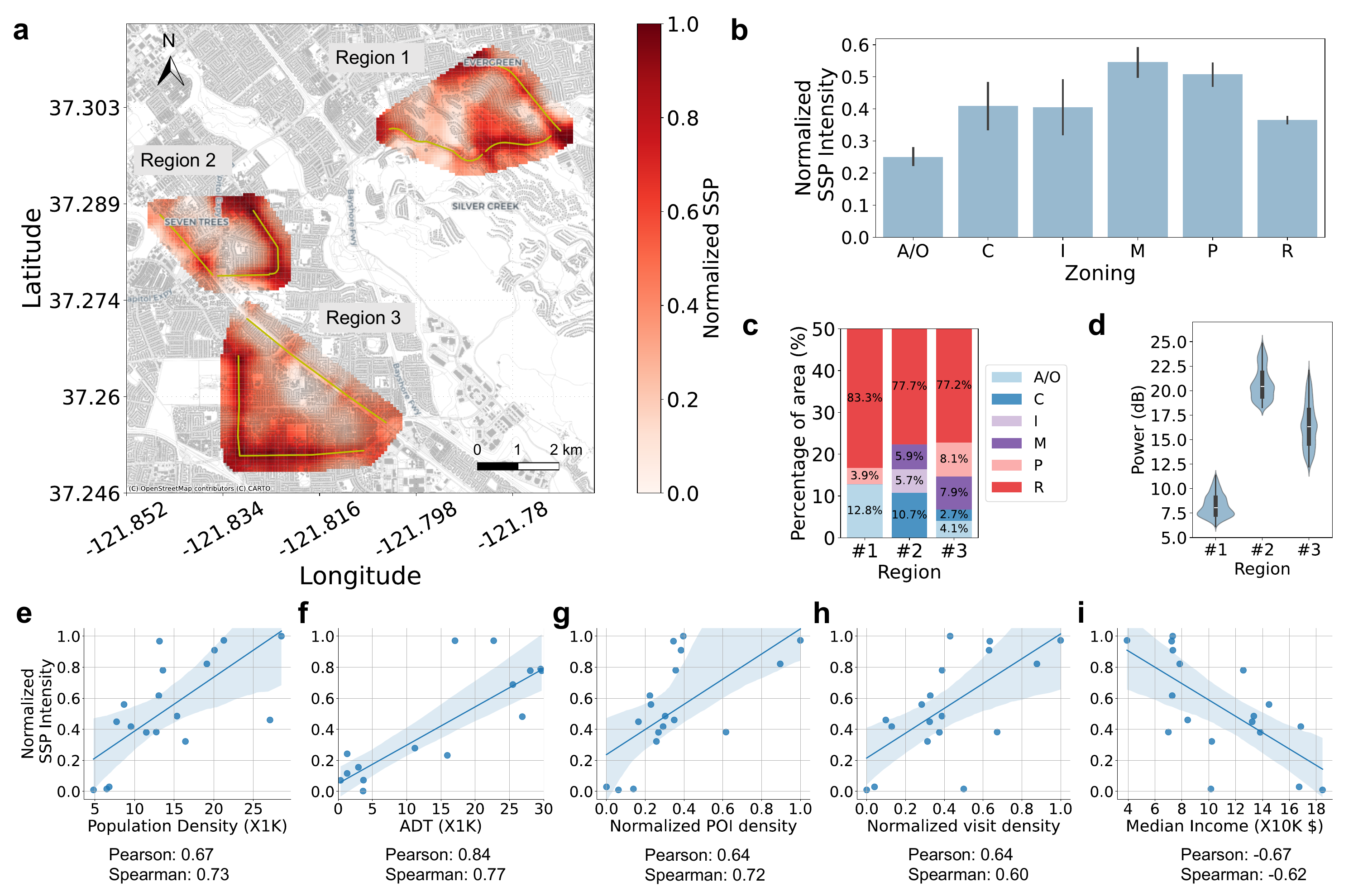}
    \caption{{\bf Seismic source power reflects the human activity intensity and land use patterns and correlates with dynamic and persistent urban features.} 
{\bf a}, SSP map averaging 6-day data from September 20th to 25th, 2023, across three regions, where a color gradient represents the normalized power within each region. 
Yellow lines mark the telecommunication cables used for DAS.
{\bf b}, Bar chart showing normalized power for each land use category: Agriculture/Open Space (A/O), Commercial (C), Industrial (I), Mixed-use (M), Public (P), and Residential (R), with error bars depicting the 95\% confidence interval. 
Due to increased urban activities, mixed-use, public, commercial, and industrial lands have higher SSP intensity. 
In contrast, areas for agricultural or open spaces exhibit the lowest SSP intensity. 
{\bf c}, The composition of land use in each region is shown by stacked bar graphs detailing percentages of six land use categories.   
{\bf d}, Violin plot illustrating the distributions of seismic power values for the three regions. 
The data shows a strong relationship between land use patterns and seismic power. 
Region 1 has the lowest power, with 12.8\% agricultural/open spaces lands. 
Regions 2 and 3, with more commercial and mixed-use lands, have higher seismic power.
{\bf e-h}, our estimated SSP intensity correlated with various persistent and dynamic urban features. 
High population density, average daily traffic (ADT), POI density, and higher visit density correlate with increased SSP intensity, where large Pearson and Spearman coefficients reveal strong positive correlations. 
{\bf i}, a negative correlation between the SSP intensity and the median income suggests higher SSP intensity in lower-income areas.
On each plot, a 95\% confidence interval is drawn using translucent bands around the solid regression line.
These results collectively suggest that human activities, population, and vehicular presence result in seismic activities, which can be a proxy for assessing urban environments and society. 
}
    \label{fig:maps}
\end{figure}

\FloatBarrier

\FloatBarrier


\section*{Supplementary material}

## Additional information

Supplementary Information is available for this paper. Correspondence and requests for materials should be addressed to Jingxiao Liu.

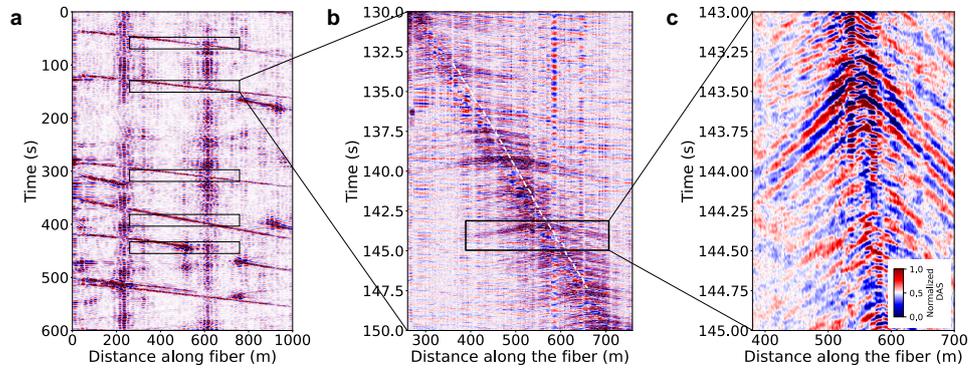

**Extended Data Fig 1 Workflow to select the vehicle-induced surface wave windows for Virtual Shot Gathers. a**, A 10-minute section of the processed DAS recordings. Black boxes indicate selected surface wave windows based on the tracked trajectories of vehicles. **b**, One of the selected time windows from 130 s to 150 s. The vehicle-induced surface waves are selected for cross-correlation to construct VSGs for the virtual shot location at $x_0 = 500\ m$. The dashed white line indicates the vehicle trajectory estimated using a specialized Kalman filter algorithm [22]. **c**, A zoomed-in view to show the surface waves generated by the moving vehicle.

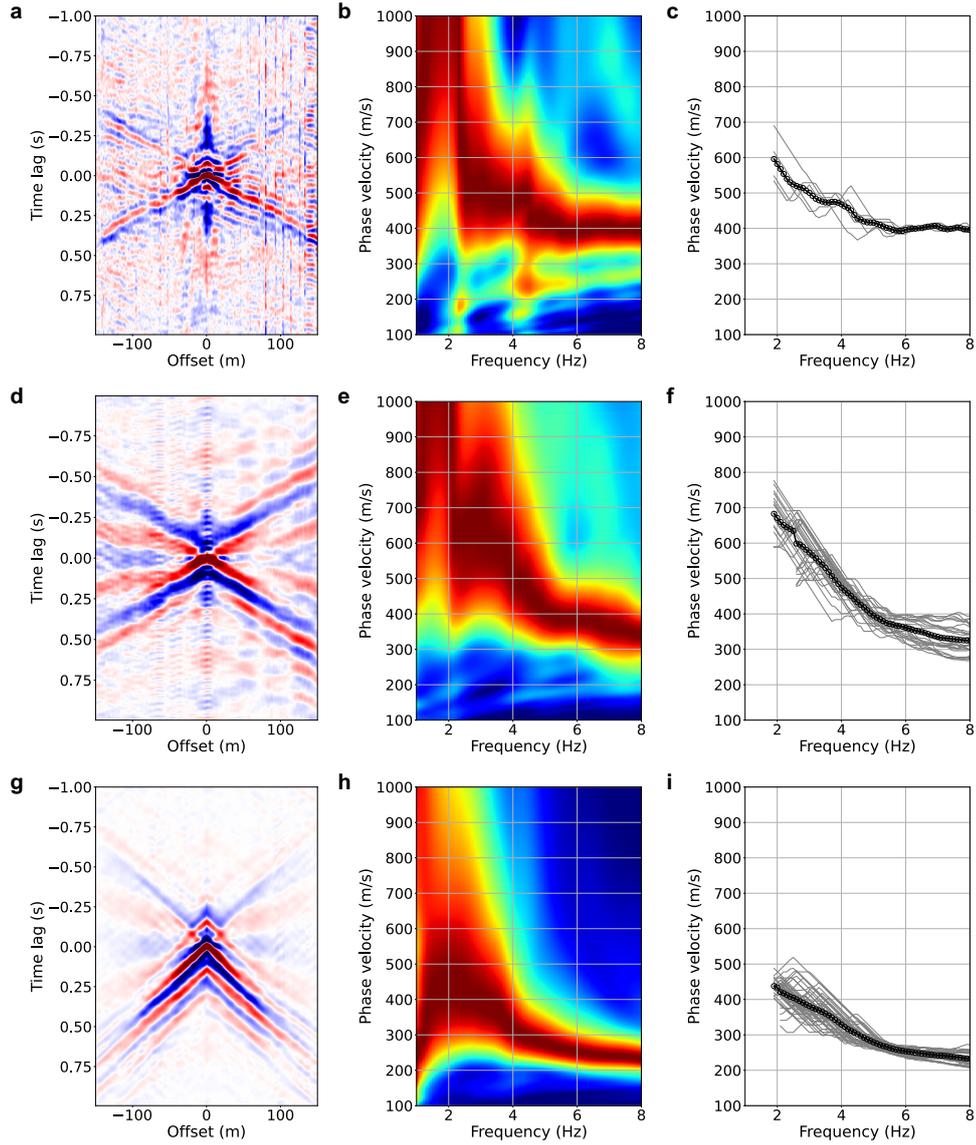

**Extended Data Fig 2 Constructed virtual short gathers (VSGs) and the estimated shear wave velocity ($v_\omega$) across the three regions. a**, The averaged VSGs in Region 1 showing the Rayleigh wave energy generated by passing vehicles. 485 vehicles are stacked. **b**, The dispersion images computed using the phase-shift method. **c**, The picked fundamental mode dispersion curve of Rayleigh surface waves from **b**. The gray lines are dispersion curves picked from VSGs at different locations/dates along the fiber in Region 1. The black line is the averaged S-wave velocity. (d)-(f) are the same as (a)-(c) for Region 2 with 2,520 vehicles stacked. (g)-(i) are the same as (a)-(c) for Region 3 with 3,492 vehicles stacked. The variation in the number of vehicles isolated for VSGs is due to differences in fiber layouts and traffic volumes across the studied regions.

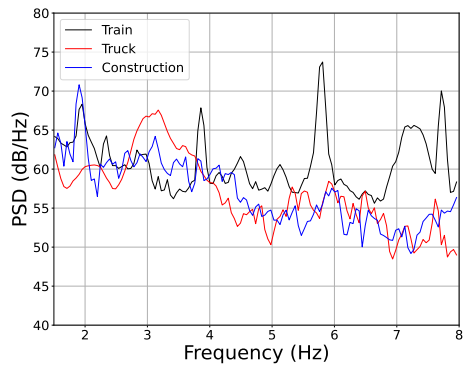

**Extended Data Fig 3 Comparison of power spectral density (PSD) profiles for three types of urban seismic sources.** This figure shows the average PSD measurements obtained from DAS channels due to three urban seismic sources: a passing train (black), a passing truck (red), and construction activities (blue). The train's signal spectrum exhibits several narrow peaks at frequencies of 2, 3.9, 5.8, and 7.7 Hz. The truck's spectrum shows a broader peak of around 3.2 Hz, and the construction activities' spectrum has a pronounced peak at approximately 1.9 Hz. Analysis of these PSD profiles can provide insights into the distinctive signatures of these urban seismic sources and facilitate the estimation of the sensing coverage in the frequency spectrum of our system (Extended Data Fig 4).

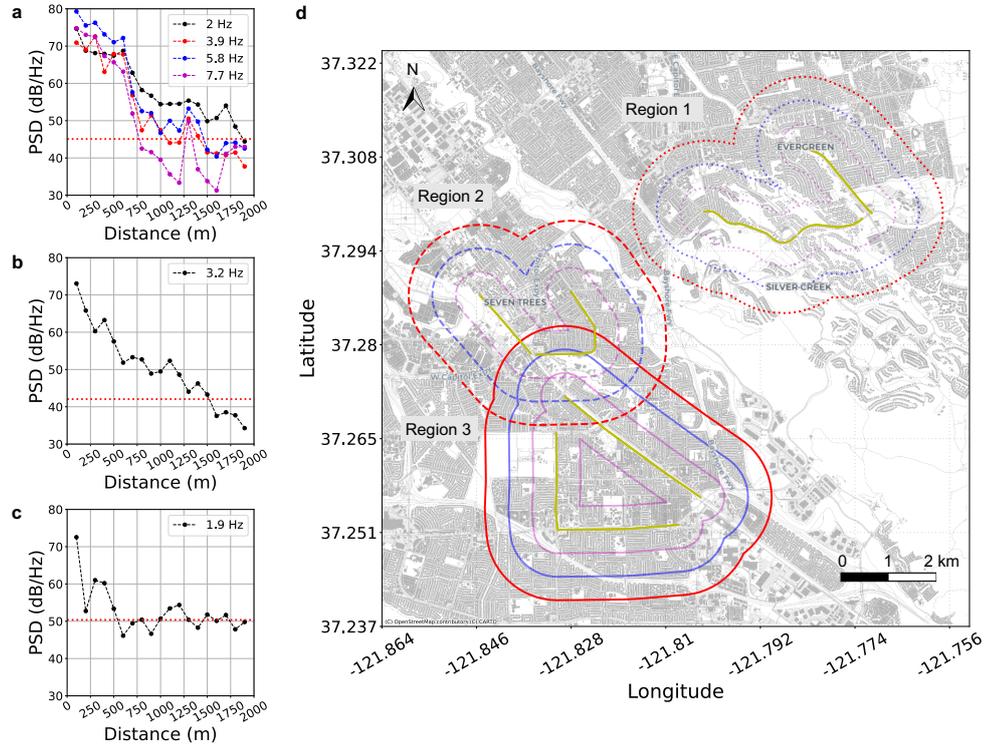

**Extended Data Fig 4 Attenuation characteristics of three urban seismic sources and a sensing coverage map of our system. a-c**, These figures show the decay of PSD values with increasing distance from urban seismic sources. **a**, The PSD values of train-induced seismic signals at frequencies of 2, 3.9, 5.8, and 7.7 Hz (dominant frequencies identified in Extended Data Fig 3) decrease as the distance from the train source increases. **b**, The PSD values of the truck-induced seismic signals in the 2.8 to 3.4 Hz frequency range decrease as the distance from the truck source increases. **c**, The figure shows the decay of PSD for construction activity-induced seismic signals at the dominant frequency of 1.9 Hz. The dotted lines in each figure indicate the background seismic noise levels averaged from noise signals associated with each seismic activity within the same timeframe. Lower-frequency seismic signals generated by passing trains exhibit less attenuation and can propagate further. The 2 Hz frequency from train sources is detectable up to 2 kilometers, while truck-induced signals persist up to about 1.5 kilometers. Construction activity generates relatively low-energy seismic signals at 1.9 Hz, detectable up to 500 meters before attenuating below background seismic noise levels. **d**, A map presenting the sensing coverage provided by our system, with contours at 500, 1000, and 1500 meters from the telecommunication cables indicated by magenta, blue, and red lines, respectively. These contours represent the sensing coverage zones for the various activities, underscoring the system's ubiquitous urban sensing capabilities.

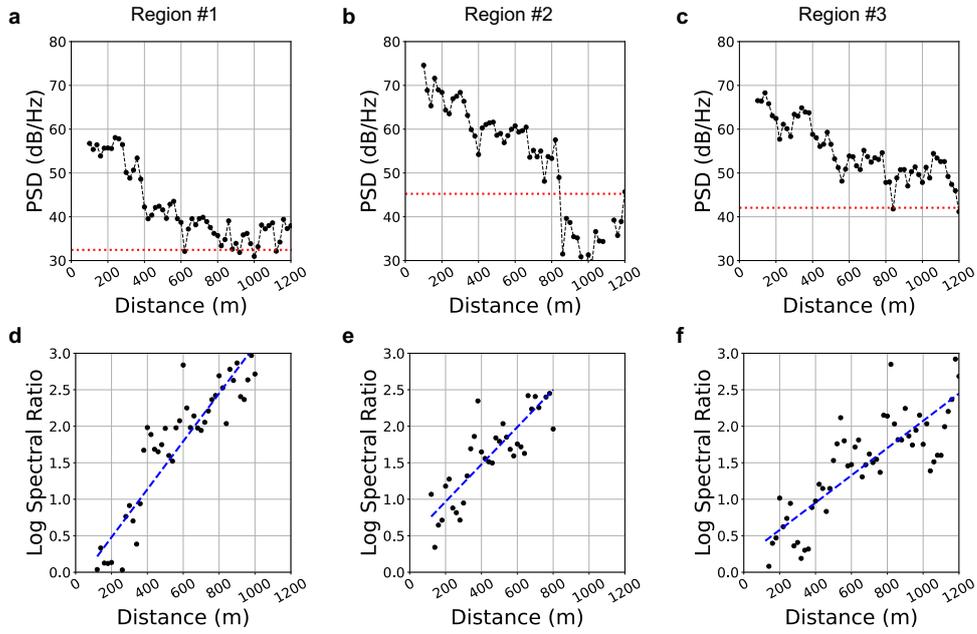

**Extended Data Fig 5 Estimation of the attenuation quality Q factors for seismic signal attenuation across three regions.** We use truck-induced seismic signals to estimate Q factors, as they are the most common seismic sources with known locations among all three regions. **a-c**, The average PSD values of the truck-induced seismic signals in the 2.8 to 3.4 Hz frequency range decrease as the distance from the truck source increases for the three regions, respectively. The red dotted lines represent the background noise levels. We calculate the logarithmic spectral ratios of the PSD values and plot the ratios at various distances from the seismic source in the scatter plots **d-f**. The dashed lines in these plots represent the least-squares fit to the data, where the slope of the line is proportional to $Q^{-1}$. The calculated Q factors for the respective regions, obtained by solving the least square problem in Eq. 7, are 6, 8, and 14 for regions 1, 2, and 3, respectively.

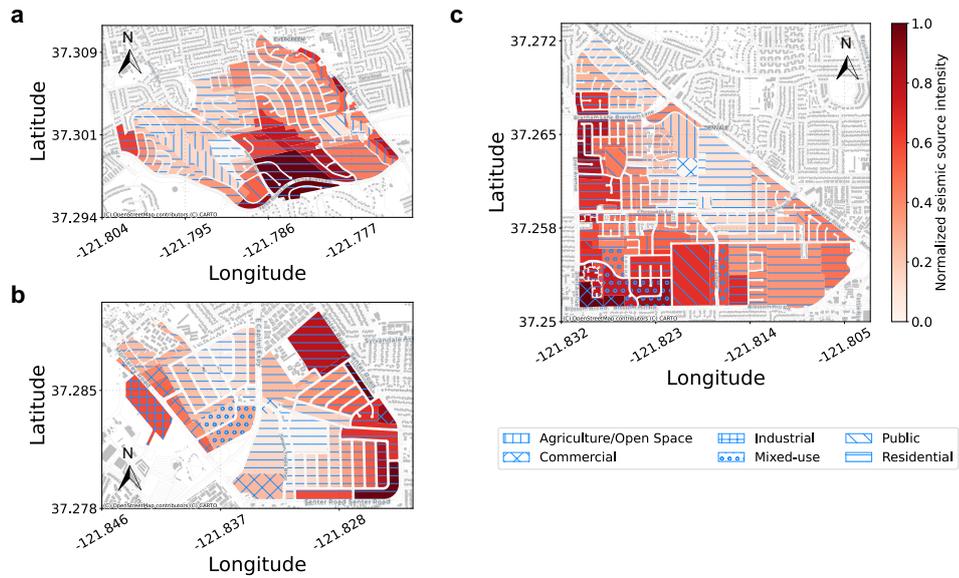

**Extended Data Fig 6 Seismic source intensity maps over a 6-day period for the three studied regions within San Jose, California. a-c**, The seismic source mapping results for regions 1, 2, and 3, respectively. The maps are color-coded to represent the normalized seismic source intensity in each census block, with the patch patterns indicating the corresponding zoning classification. These figures illustrate that areas with heightened urban activity, particularly those designated for mixed-use, public, commercial, and industrial lands, exhibit higher seismic intensity. In contrast, lands designated for agriculture or open spaces show lower seismic intensity. Additionally, areas near major roadways and highways with higher traffic volumes have increased seismic source intensity.

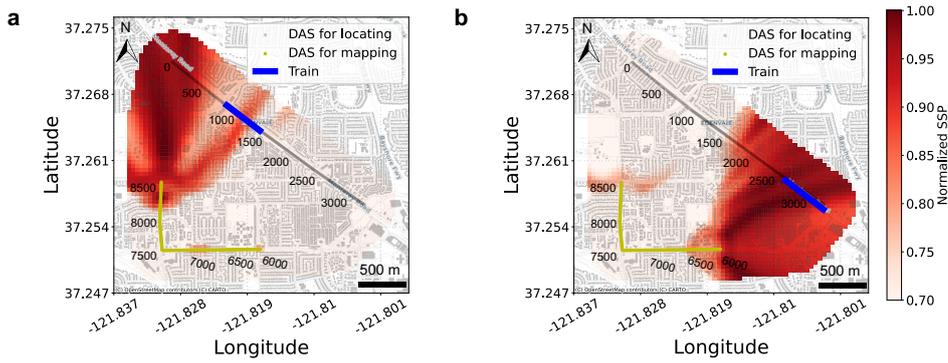

**Extended Data Fig 7 A case study to validate the seismic source mapping results for a moving freight train.** The train travels from northwest to southeast along the initial segment of our telecommunications cables (spanning channels 0 to 3,500). The quasi-static signals recorded by the adjacent DAS channels provide precise geo-locations and timings for the train's movement. Surface waves generated by the train can be detected by DAS channels up to 2 kilometers away. The yellow dots represent the active DAS channels utilized to create the SSP maps. The seismic activity has three key instances: the train's arrival (**a**), its passing through the central area of the sensing area (Fig 2b), and its departure from the monitored region (**b**). Our method effectively locates the train with areas of higher seismic power, shown as darker red on the maps, aligning closely with the ground truth positions of the train.

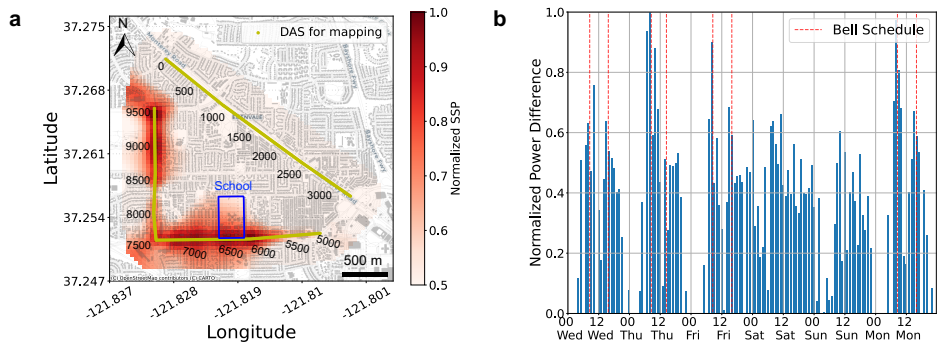

**Extended Data Fig 8 A case study of seismic source mapping of school activities and temporal seismic power differences related to school schedules. a**, SSP map from 7:40 am to 7:50 am local time in Region 1, capturing the period prior to the morning school bell. The blue box encloses the school zone, where a pronounced increase in seismic source power around the school is observed, which is attributed to increased human activities and vehicular traffic. **b**, six-day time series data of the differential power between the school zone and the nearby area within a 100-meter radius. Red vertical lines annotate the school's bell schedule, which aligns well with the large peaks of the power difference. This case demonstrates our method's capability to monitor seismic variations due to school-related activities.

41

 
\end{document}